\documentclass[prl,reprint,superscriptaddress,preprintnumbers]{revtex4-1}
\pdfoutput=1
\usepackage{amsfonts, amssymb, amsmath}
\usepackage{graphicx}
\usepackage{mathrsfs}
\usepackage{hyperref}
\usepackage[T1]{fontenc}
\usepackage{lmodern}

\usepackage{graphicx}
\usepackage[percent]{overpic}
\usepackage{caption}
\usepackage{subcaption}

\def\im{{\rm i}} 

\usepackage{color}
\definecolor{dark-gray}{gray}{0.20}
\definecolor{gray}{gray}{0.30}
\definecolor{light-gray}{gray}{0.80}
\definecolor{dark-red}{rgb}{0.7,0,0}
\definecolor{dark-green}{rgb}{0.1,0.4,0}
\definecolor{dark-blue}{rgb}{0.3,0.3,0.7}
\definecolor{light-blue}{rgb}{0.8,0.8,1}
\definecolor{blue}{rgb}{0,0,1}
\definecolor{red}{rgb}{1,0,0}
\definecolor{green}{rgb}{0,1,0}

\usepackage{hyperref}
\hypersetup{
	colorlinks=true,
	linkcolor=dark-blue,
	citecolor=dark-red,
	urlcolor=dark-green,
	linktoc=page
}

\def\Tr{{\rm Tr}\,}

\def\SL{{\rm SL}}

\def\SU{{\rm SU}}

\def\sl{\frak{sl}}

\def\su{\frak{su}}

\usepackage[paperwidth=210mm,paperheight=297mm,centering,hmargin=1.5cm,vmargin=2.5cm]{geometry}
\newcommand{\dd}{\mathrm{d}}

\newcommand{\e}{\mathrm{e}}

\newcommand{\be}{\begin{equation}}
\newcommand{\ee}{\end{equation}}
\newcommand{\bea}{\begin{eqnarray}}
\newcommand{\eea}{\end{eqnarray}}

\newcommand{\f}[2]{\frac{#1}{#2}}

\newcommand{\U}{\text{U}}

\begin{document}

\title{Janus and J-fold Solutions from Sasaki-Einstein Manifolds}

\date{\today}

\author{Nikolay Bobev}

\affiliation{Instituut voor Theoretische Fysica, KU Leuven, Celestijnenlaan 200D, B-3001 Leuven, Belgium}

\author{Fri{\dh}rik Freyr Gautason}

\affiliation{Instituut voor Theoretische Fysica, KU Leuven, Celestijnenlaan 200D, B-3001 Leuven, Belgium}

\author{Krzysztof Pilch}

\affiliation{Department of Physics and Astronomy, University of Southern California, Los Angeles, CA 90089, USA}

\author{Minwoo Suh}

\affiliation{Department of Physics, Kyungpook National University, Daegu, 41566, Korea}

\author{Jesse van Muiden}

\affiliation{Instituut voor Theoretische Fysica, KU Leuven, Celestijnenlaan 200D, B-3001 Leuven, Belgium}

\begin{abstract}
\noindent We show that for every Sasaki-Einstein manifold, $M_5$, the AdS$_5\times M_5$ background of type IIB supergravity admits two universal deformations leading to supersymmetric AdS$_4$ solutions. One class of solutions describes an AdS$_4$ domain wall in AdS$_5$ and is dual to a Janus configuration with $\mathcal{N}=1$ supersymmetry. The other class of backgrounds is of the form AdS$_4\times S^1\times M_5$ with a non-trivial $\SL(2,\mathbb{Z})$ monodromy for the IIB axio-dilaton along the $S^1$. These AdS$_4$ solutions are dual to three-dimensional $\mathcal{N}=1$ SCFTs. Using holography we express the $S^3$ free energy of these theories in terms of the conformal anomaly of the four-dimensional $\mathcal{N}=1$ SCFT arising from D3-branes on the Calabi-Yau cone over $M_5$. 

\end{abstract}

\pacs{}
\keywords{}

\maketitle

\section{Introduction}\label{sec:intro}
%

Defects and interfaces play an important role in the dynamics of quantum field theory  and find many applications ranging from condensed matter physics to string theory. Their physics is often strongly coupled and thus difficult to study with conventional techniques. It is therefore natural to use AdS/CFT to study properties of defects and interfaces in strongly interacting QFTs. The best understood examples of the holographic correspondence arise from string or M-theory with some amount of unbroken supersymmetry. Indeed, the codimension one interfaces and defects have been studied extensively in the context of the duality between IIB string theory on AdS$_5\times S^5$ and $\mathcal{N}=4$ SYM.

A particular class of interfaces of interest to us here are the so called Janus interfaces \cite{Bak:2003jk,Clark:2004sb}. In $\mathcal{N}=4$ SYM they arise from studying the theory with a position dependent gauge coupling along one of the spatial directions on $\mathbb{R}^{1,3}$. By choosing a specific position dependence of the coupling and turning on additional operators in the $\mathcal{N}=4$ theory, the interface can preserve three-dimensional superconformal symmetry. For Janus interfaces with 3d $\mathcal{N}=4$ supersymmetry this set-up was studied in detail in \cite{DHoker:2006qeo,Gaiotto:2008sd,Gaiotto:2008ak}. S-duality and a non-trivial profile for the $\theta$ angle of the $\mathcal{N}=4$ theory imply the existence of new strongly coupled 3d $\mathcal{N}=4$ SCFTs localized on the interface. These so-called $T[U(N)]$ theories serve as strongly coupled building blocks which can be used to construct 3d ${\cal N}=4$ QFTs. In particular, it was shown in \cite{Assel:2018vtq}, see also \cite{Ganor:2014pha},  that one can gauge the $\U(N)\times\U(N)$ global symmetry and introduce Chern-Simons interaction terms to arrive at new three-dimensional $\mathcal{N}=4$ SCFTs. We refer to this type of construction as J-fold. 

These Janus and J-fold constructions in $\mathcal{N}=4$ SYM have a natural realization in type IIB supergravity. The Janus solutions are realized as deformations of AdS$_5\times S^5$ to a domain-wall with AdS$_4$ slicing, two asymptotic AdS$_5$ regions and a squashed metric on $S^5$ \cite{DHoker:2007zhm}. The J-fold configuration is holographically dual to an AdS$_4\times S^1\times S^5$ solution with a non-trivial profile for the IIB axio-dilaton along the $S^1$ \cite{Inverso:2016eet,Assel:2018vtq}.

\begin{figure}
\centering
\begin{overpic}[scale=0.34]{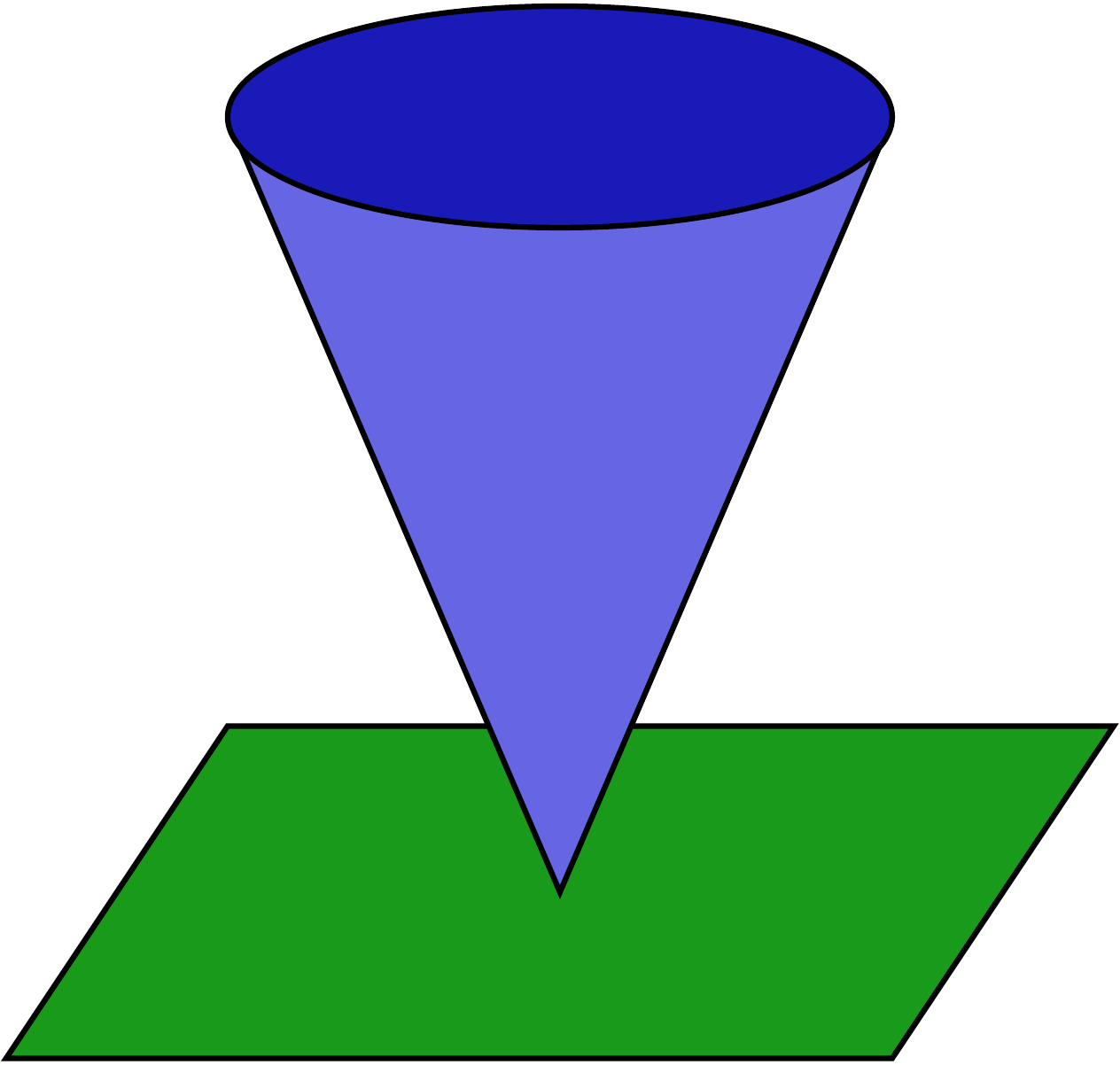}
\put(43.5,80.5){\color{white}{\large SE$_5$}}
\end{overpic} 
\caption{4d $\mathcal{N}=1$ SCFT realized on $N$ coincident D3-branes probing a CY cone over a SE manifold.}
\label{cone}
\end{figure}
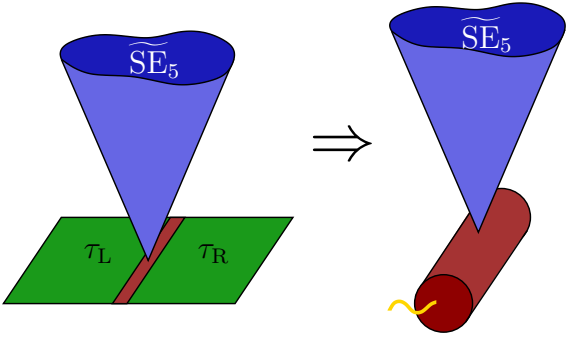
\begin{figure}
\begin{minipage}[c][][t]{.28\textwidth}
\begin{overpic}[scale=0.3]{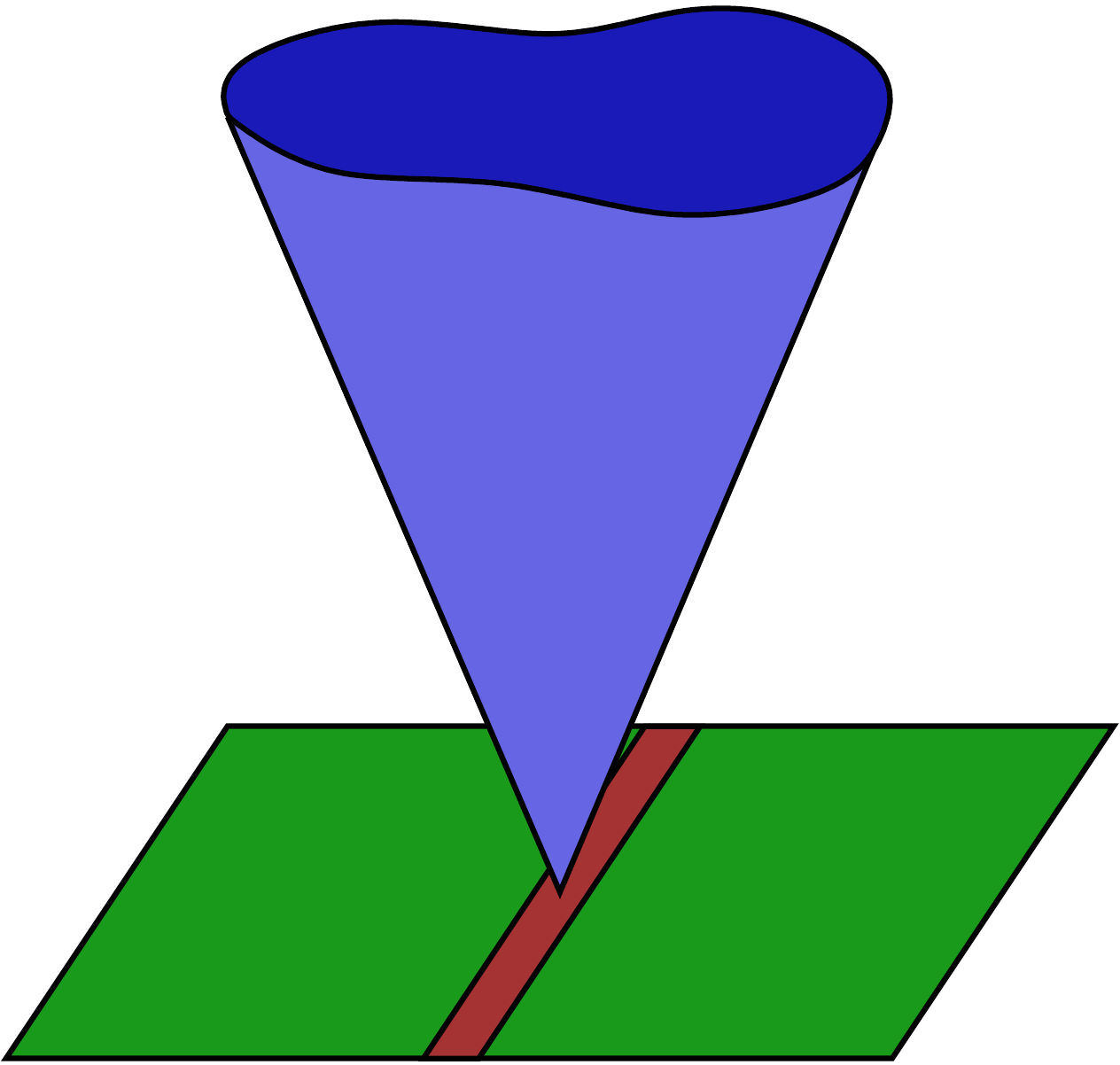}
\put(43.5,80){\color{white}{\large $\widetilde{\text{SE}}_5$}}
\put(28,16.3){{\large$\tau_{\rm L}$}}
\put(67,16.3){{\large$\tau_{\rm R}$}}
\put(105,50){{\Huge$\Rightarrow$}}
\end{overpic}
\end{minipage}%
\begin{minipage}[c][][t]{.2\textwidth}
\begin{overpic}[scale=0.3]{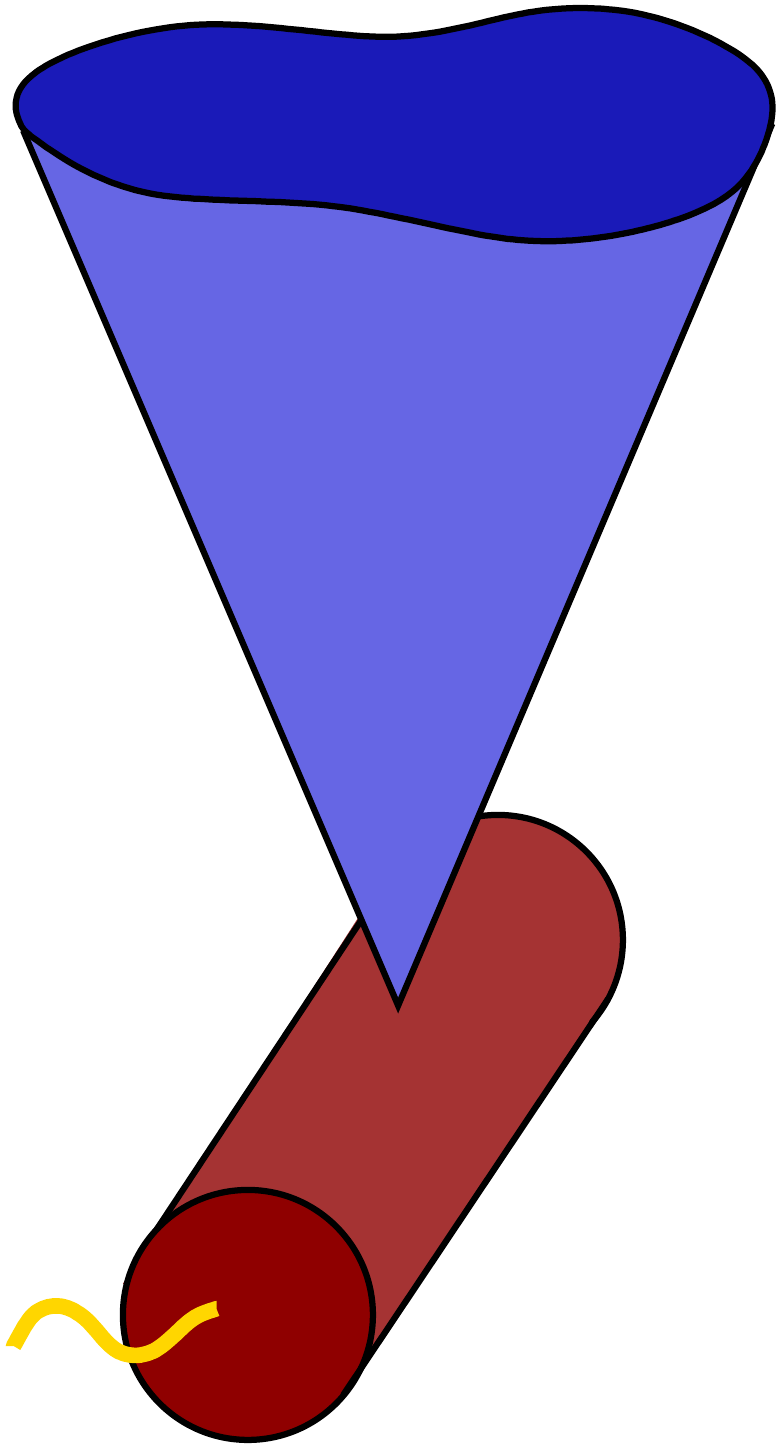}
\put(22.5,86){\color{white}{\large $\widetilde{\text{SE}}_5$}}
\end{overpic}
\end{minipage}
\caption{Left: A Janus configuration for the D3-brane worldvolume theory in Figure~\ref{cone} with gauge couplings $\tau_{\rm L}$ and $\tau_{\rm R}$ on the left and right side of the interface. \\Right: The J-fold configuration associated with the same SE manifold.}
\label{squashedcone}
\end{figure}

Our goal here is to use IIB supergravity to generalize the Janus and J-fold constructions to four-dimensional SCFTs where the theory on the interface preserves three-dimensional $\mathcal{N}=1$ supersymmetry. A natural starting point is to consider deformations of the AdS$_5\times M_5$ backgrounds of type IIB supergravity where $M_5$ is a Sasaki-Einstein (SE) manifold. There are infinite classes of such manifolds with explicit metrics known as $Y^{p,q}$ \cite{Gauntlett:2004yd} and $L^{a,b,c}$ \cite{Cvetic:2005ft,Martelli:2005wy}. The 4d $\mathcal{N}=1$ quiver gauge theories dual to these supergravity solutions are  well-understood \cite{Benvenuti:2004dy,Franco:2005sm,Benvenuti:2005ja,Butti:2005sw} and arise from the dynamics of D3-branes probing the tip of the Calabi-Yau cone over $M_5$, see Figure~\ref{cone}. The deformations of this D3-brane setup we study are illustrated in Figure~\ref{squashedcone}.

To construct the $\mathcal{N}=1$ AdS$_4$ solutions describing Janus and J-fold configurations, we make use of the fact that, for every SE manifold, IIB supergravity admits a consistent truncation to five-dimensional $\mathcal{N}=2$ gauged supergravity coupled to one hypermultiplet. This is a subtruncation of the more general truncation of IIB supergravity on SE manifolds studied in \cite{Cassani:2010uw,Skenderis:2010vz,Gauntlett:2010vu}, see also \cite{Gubser:2009qm,Liu:2010pq} for some related results. We can thus first construct Janus and J-fold solutions in five dimensions by solving a system of coupled nonlinear differential equations and then uplift the result to IIB supergravity.

We note that for the special case when $M_5$ is the round $S^5$ the Janus solution was found in \cite{Clark:2005te,DHoker:2006vfr,Suh:2011xc} and the J-fold solution was very recently presented in \cite{Guarino:2019oct}. The local form of the J-fold solutions for general SE $M_5$ was also presented in \cite{Lust:2009mb}. Here we show how to make this solution globally well-defined by imposing an appropriate $\SL(2,\mathbb{Z})$ monodromy, characterized by an integer, $n$, along the $S^1$ direction. We also derive an universal relation, valid in the planar limit, between the $S^3$ free energy of the 3d $\mathcal{N}=1$ SCFT dual to the J-fold solution and the conformal anomaly of the 4d $\mathcal{N}=1$ SCFT dual to the original AdS$_5\times M_5$ solution.

%
%
%


\section{Five-dimensions}\label{sec:5d}

The solutions we consider can be described within five-dimensional ${\cal N}=2$ gauged supergravity coupled to one hypermultiplet. All solutions of this theory can be uplifted to type IIB supergravity as we review below.

\begin{figure}
\centering
\includegraphics[width=0.38\textwidth]{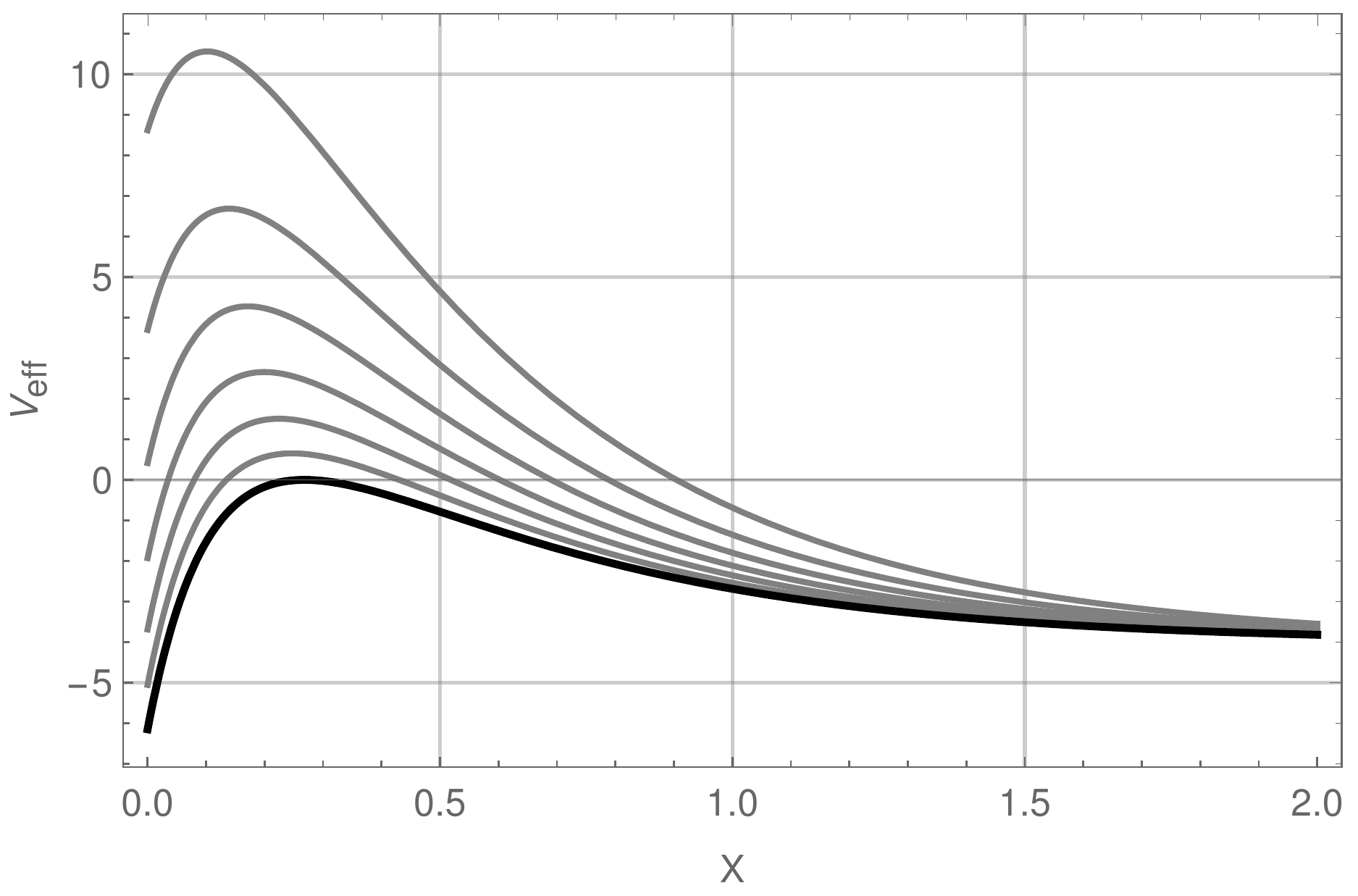}
\caption{\label{effpot}A plot of the effective potential \eqref{SU3classicalmech} in the allowed range $0\le{\cal I}<1$. The black curve has ${\cal I}=1$.}
\end{figure}

The five-dimensional theory consists of a metric, two gravitini, and an $\U(1)$ gauge field, which together form the ${\cal N}=2$ gravity multiplet, in addition to two spin-1/2 fermions and four scalars forming the hypermultiplet. Here, we consider solutions of the $\mathcal N = 2$ theory for which both the gauge field and the fermions are set to zero. This is a consistent truncation at the level of equations of motion. The bosonic Lagrangian of this truncated subsector is
\be\label{Eq:5Dlagrangian}
\mathcal{L} = \frac{\sqrt{|g_{5}|}}{16\pi G_N}\left(R_{5} +\f14 \Tr\big[\partial_\mu M \partial^\mu M^{-1}\big] - \mathcal{P}\right)\,,
\ee
where ${\cal P}$ is the potential on the scalar manifold 
\be
{\cal M} = \f{\SU(2,1)}{\U(2)}\,,
\ee
parametrized by the four hypermultiplet scalars through the sigma model matrix $M$. We will parametrize this manifold in a non-standard but convenient way. We start by defining the four non-compact generators of $\su(2,1)$,
\be
[\mathfrak{e}^a]_{ij} =\delta_{i}^{a}\delta_j^3+\delta_j^a\delta_i^3~,\qquad [\mathfrak{f}^a]_{ij} = \im\epsilon^a_{\phantom{a}ij}\,,
\ee
where $a=1,2$ and $i,j=1,2,3$. Together with the compact generators $\mathfrak{h}^1 = [\mathfrak{e}^1,\mathfrak{f}^1]$ and $\mathfrak{h}^2 = [\mathfrak{e}^2,\mathfrak{f}^2]$, these generators form two copies of $\su(1,1)\simeq\sl(2,{\bf R})$. Note that these two $\su(1,1)$'s do not commute. Now the scalar matrix is given by
\be\label{Mmatrixsu21}
M = U^\dagger U\,,\qquad U = \e^{\chi \mathfrak{e}^1}\cdot \e^{\f{2\omega+c}{4} \mathfrak{h}^2}\cdot \e^{\varphi \mathfrak{e^2}}\cdot \e^{-\f{c}{4} \mathfrak{h}^2}\,.
\ee
In this parametrization the scalar kinetic terms take the explicit form
\be
\begin{split}
{\cal L}_\text{kin} &= \f14 \Tr\big[\partial_\mu M \partial^\mu M^{-1}\big]\\
&=   -2(\partial \chi)^2 -\tfrac{1}{2} \sinh^22\chi (\partial\omega - \sinh^2\varphi~\partial c)^2 \\
&\quad -\tfrac{1}{2} \cosh^2\chi\big[4(\partial \varphi)^2 + \sinh^22\varphi(\partial c)^2\big]~.
\end{split}
\ee
The potential can be written in terms of a superpotential
\be
\mathcal P = \frac{1}{2}\left[ (\partial_\chi W)^2 - \frac{8}{3}W^2\right]\,,\quad W =-\f{3g}{2}\cosh^2\chi\,.
\ee
This theory enjoys an exact $\SL(2,\mathbb{R})_S$ symmetry that will play an important role in the following. This symmetry is a direct consequence of the $\SL(2,\mathbb{R})$ symmetry of type IIB supergravity. In five dimensions, the $\SL(2,\mathbb{R})_S$ in question is generated by $\{\mathfrak{e}^2,\,\mathfrak{f}^2,\,\mathfrak{h}^2\}$. It acts on the scalars as 
\be\label{SL2action}
M\mapsto R^\dagger M R~,
\ee
and does not act on the five-dimensional metric $g_{5}$. We are interested in solutions for which the metric takes the form
\be\label{5Dmetric}
\dd s_5^2 = \dd r^2 + \e^{2A(r)} \dd s_{\text{AdS}_4}^2\,,
\ee
and the scalars are functions only of $r$. The complete set of BPS equations for this Ansatz can be derived in a straightforward manner \cite{BGPSvM}, see also \cite{Clark:2005te,Suh:2011xc}. The spin-1/2 supersymmetry variations lead to the following three equations:
\be\label{eq:bps12}
\begin{split}
(\chi')^2 &=\, \tfrac{1}{4} (\partial_\chi W)^2 -\cosh^2\chi\sec^2(c+2\omega)(\varphi')^2\,,\\
\omega' &=\, \sinh^2\varphi\,(c')\,,\\
\sinh2\varphi&\,(c') =\, -2\tan(c+2\omega)(\varphi')\,,
\end{split}
\ee
where the prime denotes a derivative with respect to $r$. The spin-3/2 supersymmetry variations yield
\be
\begin{split}\label{eq:bps32}
A' =&\, -\tfrac{1}{3} \coth\chi\, (\chi')\,,\\
\varphi' =&\, 3\e^{-A}\cos(c+2\omega)\text{sech}\chi\tanh\chi\,.
\end{split}
\ee
To solve this system of equations we first notice that the equations for $c$ and $\omega$ can be solved directly in terms of the scalar $\varphi$
\be\label{phasesol}
\begin{split}
\sin(c +2 \omega) =&\, \f{{\cal J}}{\sinh2\varphi}\,,\\
\cos^2(c - c_0) =&\, \f{\sinh^22\varphi-{\cal J}^2}{\sinh^22\varphi(1+{\cal J}^2)}\,,
\end{split}
\ee
where we have introduced two integration constants ${\cal J}$ and $c_0$. Similarly we can integrate for $A$ in \eqref{eq:bps32}
\be\label{eq:Asol}
\e^{-6A} = \left(\f{3 g }{5}\right)^6\f{5\sinh^2\chi}{{\cal I}^3}\,,
\ee
where ${\cal I}$ is another integration constant. We are now left with solving for the scalars $\varphi$ and $\chi$. In order to simplify the remaining expressions, we define a shifted metric function $\e^{-3X} = \sinh\chi$. Using the BPS equations and \eqref{eq:Asol}, one finds that $X$ satisfies 
\be\label{SU3classicalmech}
\begin{split}
&\f{4}{g^2}(X')^2 + V_\text{eff} = 0\,,\\
&V_\text{eff} = 16\e^{-2X}\left(\f{9}{5^{5/3}{\cal I}}- \e^{-4X}\cosh^2(3X)\right)\,. 
\end{split}
\ee
This reduces the problem of finding $X$ to that of a classical particle with zero energy scattering off the potential $V_\text{eff}$. As $X$ tends to $\pm\infty$ the metric function $A$ diverges while the scalar $\chi\to0$. In this limit we recover the AdS$_5$ vacuum. In order to get a regular Janus solution with two asymptotic AdS$_5$ regions, the integration constant ${\cal I}$ must lie in the range $0\le{\cal I}\le 1$. Only in this range $V_\text{eff}$ has a maximum on the positive real axis for which $V_\text{eff}\ge 0$, see Figure~\ref{effpot}.
The dilaton $\varphi$ can then be written as
\be\label{SU3dilatonsol}
\cosh2\varphi =\cosh2F+\f12\e^{-2F}{\cal J}^2\,,
\ee
where 
\be
F = F_0 \pm \int \f{9 \e^{-X} \dd X }{\cosh(3X)\sqrt{-5^{5/3}{\cal I}V_\text{eff}}}\,.
\ee
%
\begin{figure}
\centering
\includegraphics[width=0.38\textwidth]{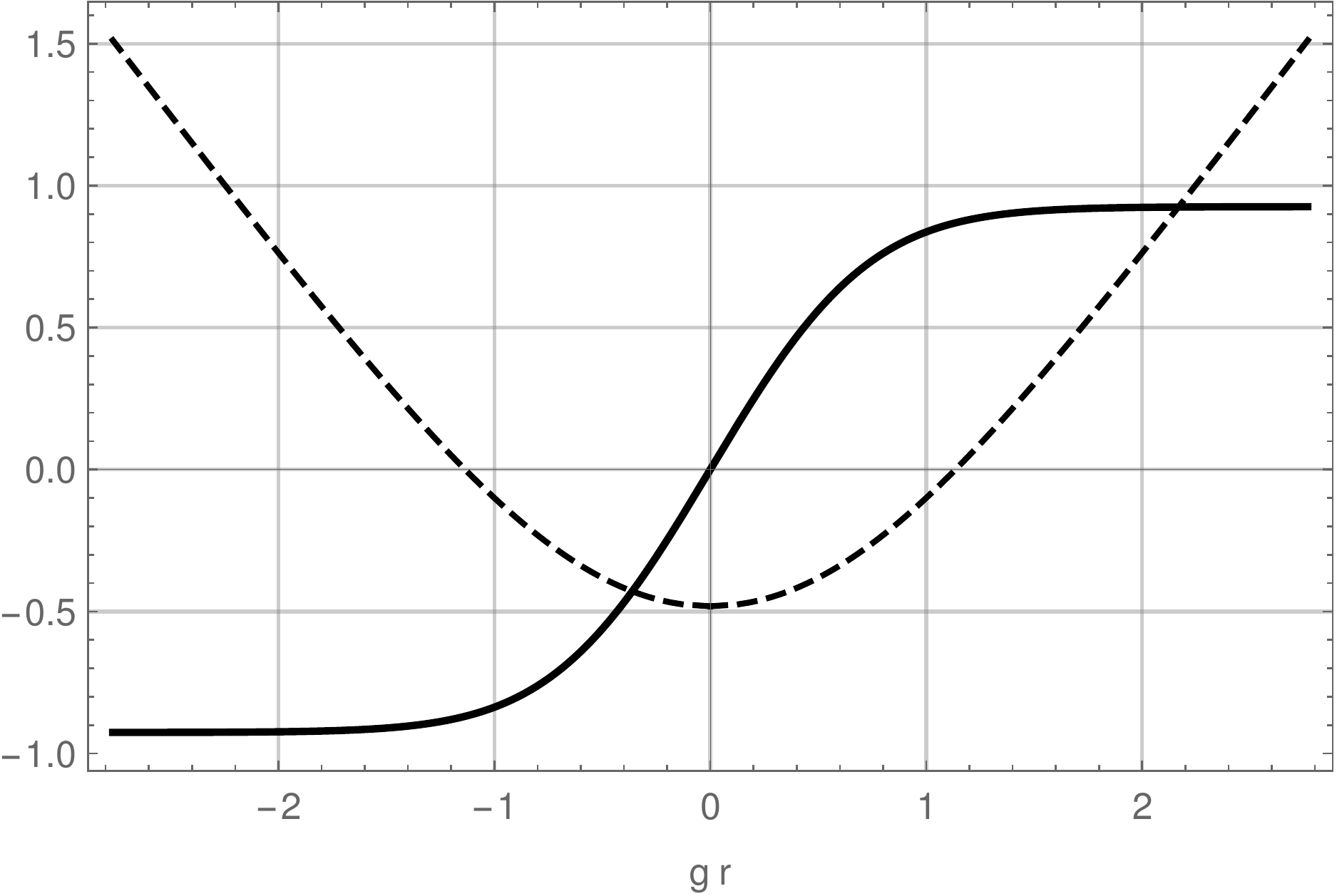}
\caption{\label{SU3plot}A plot of the function $4(F(r)-F_0)$ (solid curve) for ${\cal I}=4/5$. We also display the function $X-1$ (dashed curve), which determines the metric function $A$.}
\end{figure}
Using the classical mechanics problem and the integral expression for $F$ we can perform two numerical integrations to completely solve the system. The scalar fields $\varphi$, $c$, and $\omega$ are then obtained from \eqref{SU3dilatonsol} and \eqref{phasesol}. In Figure~\ref{SU3plot} we display a sample numerical solution. 

Now let us focus on ${\cal I}=1$. Here we can again construct Janus solution for which the scalar $X$ comes in from $+\infty$ and scatters off the potential. However, now the critical point of the effective potential has exactly zero energy, see Figure~\ref{effpot}. This implies that the particle can stay as long as we wish at the critical point located at $6X=\log 5$ before returning back to $X=\infty$.   In fact it can stay there indefinitely, i.e.~there are exact solutions to the BPS equations for which $X$ is constant. These solutions do not have asymptotic regions where the metric approaches the AdS$_5$ vacuum. Instead the metric is simply of the product form AdS$_4\times \mathbb{R}$
\be
\dd s_5^2 = \f{5}{9g^2}\left(4\dd \rho^2 + 5\dd s_{\text{AdS}_4}^2\right)\,,
\ee
where we have changed to the coordinate $\rho = 3g r/(2\sqrt{5})$. Although $\sinh^2\chi = 1/5$ is constant, the remaining scalar fields are non-trivial functions of $\rho$ and  are determined by the function $F$, see \eqref{phasesol} and \eqref{SU3dilatonsol},  which is given by
\be
F = F_0 + \rho ~.
\ee
A particular solution of this type can be compactified to an S-fold solution. An S-fold is a solution of type IIB string theory that is periodic up to an $\SL(2,\mathbb{Z})$ transformation of the fields. The S-folds we construct are closely related to the Janus solutions above and we refer to them as J-folds. In five dimensions this compactification is achieved by periodically identifying the $\rho$ coordinate $\rho \sim \rho+\rho_0$ while making sure that that all fields are periodic up to an $\SL(2,\mathbb{R})_S$ transformation. In order to obtain a physical background of type IIB string theory we must act with an element of $\SL(2,\mathbb{Z})_S$. The hyperbolic element we consider is the same as the one used in \cite{Inverso:2016eet,Assel:2018vtq} \footnote{More general hyperbolic elements of $\SL(2,\mathbb{Z})$ were considered in \cite{Inverso:2016eet,Assel:2018vtq} which could also be incorporated into our setup.}
\be\label{Jmatrix}
\mathfrak{J}_n  = \begin{bmatrix} n & 1\\-1&0\end{bmatrix}\,.
\ee
This transformation acts on the scalar matrix $M$ as in \eqref{SL2action} where $\mathfrak{J}_n$ is considered as an element of $\SL(2,\mathbb{R})_S \subset \SU(2,1)$. It turns out that both $\chi$ and $c+2\omega$ do not transform under the action of $\SL(2,\mathbb{R})_S$ and so they must be periodic functions of $\rho$ for a consistent compactification. For the solution in question this is only possible when these scalars are constant. The condition $\chi=\,$constant is already implied by setting $\mathcal{I}=1$. Setting $c+2\omega$ to be constant implies that ${\cal J}=0$ or $c+2\omega=0$. In order to properly compactify the solution into a J-fold using \eqref{Jmatrix} we must take $c = \pi/2$ and perform a global $\SL(2,\mathbb{R})_S$ rotation of the scalar matrix such that $U$ in \eqref{Mmatrixsu21}  takes the form
\be
U = \e^{\chi \mathfrak{e}^1}\cdot \e^{(\varphi_0+\rho_0/2+\rho) \mathfrak{e^2}}\cdot \e^{-\f{\pi}{8} \mathfrak{h}^2}\cdot \e^{\f12 \log\coth(\rho_0/2)\mathfrak{e}^2}\,,
\ee
where $\sinh^2 \chi = 1/5$ and we use $\varphi_0+\rho_0/2$ instead of the constant $F_0$. This solution has the desired property
\be
\mathfrak{J}_n^\dagger M(\rho+\rho_0) \mathfrak{J}_n = M(\rho)\,,\qquad n = 2\cosh\rho_0\,.
\ee
The identification of $n$ with $\rho_0$ implies that the period of our compactified coordinate is quantized, $n\in\mathbb{Z}$ and $n>2$. We have verified that for this solution the supersymmetry transformation parameters do not depend on the coordinate $\rho$ and they do not transform under the $\SL(2,\mathbb{Z})$ transformation \eqref{Jmatrix}. We therefore conclude that the J-fold preserves ${\cal N}=1$ supersymmetry.

\section{Uplift to IIB}\label{sec:10d}

The five-dimensional model above uplifts to the following IIB background. The metric in an Einstein frame is \footnote{We use the same type IIB supergravity conventions as in \cite{Bobev:2018hbq} which agree with the ones in \cite{Polchinski:1998rr}.}
\begin{equation}\label{eq:10dmetgen}
\dd s^2_{10} = \cosh\chi\,\dd s^2_5 + \frac{4}{g^2}\Big(\f{\dd s^2_{4}}{ \cosh\chi}+\cosh\chi~ \zeta^2\Big)\,.
\end{equation}
Here $\dd s^2_{4}$ is a local K\"ahler-Einstein metric with K\"ahler form, $J$, which satisfies $2J=\dd\sigma$, and $\zeta=\dd\phi+\sigma$. To present the 2-form fluxes in a compact form we make use of the holomorphic $(2,0)$-form, $\Omega$, on the K\"ahler-Einstein base which satisfies the identities (see \cite{Sparks:2010sn} for a review on Sasaki-Einstein geometry)
\begin{equation}
\Omega \wedge \overline\Omega = 2J\wedge J\,, \qquad \dd \Omega = 3\im \sigma \wedge \Omega\,.
\end{equation}
The two-form potential can be written as 
\begin{equation}\label{eq:C2B2}
C_2 - \tau B_2 = -\frac{4 \im}{g^2}\frac{\e^{-\im\omega}\tanh\chi}{\cosh\varphi+\im \e^{\im c}\sinh\varphi}e^{3\im\phi}\Omega\,.
\end{equation}
The type IIB four-form is
\begin{equation}\label{eq:dC4}
 C_4  = \frac{16}{g^4}\dd\phi \wedge \sigma\wedge J \,,
\end{equation}
and the axio-dilaton is given by
\begin{equation}
\tau = C_0 + \im \e^{-\Phi} = \frac{\sinh2\varphi\cos c+\im}{\cosh2\varphi-\sinh2\varphi\sin c}\,.
\end{equation}
The BPS equations in \eqref{eq:bps12} and \eqref{eq:bps32} imply the equations of motion of IIB supergravity. This is expected on general grounds based on the consistent truncation results in \cite{Cassani:2010uw,Skenderis:2010vz,Gauntlett:2010vu}. In our context this implies that the Janus and J-fold solutions discussed in the previous section lead to a supersymmetric solution of IIB supergravity for any five-dimensional SE manifold.

Note that the two-forms $C_2$ and $B_2$ in \eqref{eq:C2B2} have an explicit dependence on $\phi$ and thus the Reeb vector isometry of the Sasaki-Einstein manifold is not a symmetry of the IIB background. If the metric $\dd s^2_4$ in \eqref{eq:10dmetgen} on the K\"ahler-Einstein base has any isometries, they are preserved by the fluxes and the axio-dilaton.

\section{J-folds}\label{sec:Jfolds}
Since we performed a global $\SL(2,\mathbb{R})$ transformation to obtain the J-fold solution, we must separately uplift that solution to type IIB supergravity. The IIB J-fold solution has the metric
\begin{equation}
\dd s^2_{10} = \sqrt{\frac{5}{6}}\frac{2}{3g^2}\left(4\dd \rho^2 + 5\dd s_{\text{AdS}_4}^2 + 6\dd s^2_{4}+ \tfrac{36}{5}\zeta^2\right)\,.
\end{equation}
The axio-dilaton is 
\begin{equation}
\tau = C_0 + \im \e^{-\Phi} = \frac{\cosh(2\varphi+\rho_0)+\im \sinh\rho_0}{\cosh2\varphi}\,.
\end{equation}
The two-form is
\begin{equation}
C_2 - \tau B_2 = -\frac{2 \im}{g^2}\frac{\sqrt{\tfrac{2}{3}\sinh\rho_0}}{\cosh\varphi+\im \sinh\varphi}e^{ 3\im\phi}\Omega\,.
\end{equation}
The four-form is the same as in \eqref{eq:dC4}. The only non-trivial function in the solution above is
\begin{equation}
\varphi = \varphi_0+\rho\,,
\end{equation}
where $\varphi_0$ is an integration constant. The coordinate $\rho$ is periodic with the identification $\rho \sim \rho+\rho_0$. Consistency with the $\SL(2,\mathbb{Z})$ symmetry of IIB string theory imposes the constraint \footnote{We note that in ten dimensions the periodic identification is accompanied with the $\SL(2,\mathbb{Z})$ action $(\mathfrak{J}^{-1}_n)^{\rm T}$.}
\begin{equation}\label{eq:nIIB}
n = 2 \cosh(\rho_0) \,, \qquad n \in \mathbb{Z}\,, ~~ n>2\,.
\end{equation}
We note in passing that similar J-fold solutions were constructed in \cite{Robb:1984uj}, however, those solutions are non-supersymmetric and it is unclear whether they are perturbatively stable.

Equipped with these $\mathcal{N}=1$ AdS$_4$ solutions of IIB string theory with a compact internal space it is natural to conjecture that for each such solution there is a dual 3d $\mathcal{N}=1$ SCFT. The $S^3$ free energy of this SCFT, in the planar limit, can be computed from the supergravity solution above using the standard AdS/CFT dictionary and reads
\be\label{eq:FS3}
{\cal F}_{S^3} = \sqrt{\f{5^5}{3^6}}  \,\text{arccosh}(n/2) a_{4d}\,.
\ee
Here $a_{4d}$ is the central charge of the 4d $\mathcal{N}=1$ SCFT associated with the Sasaki-Einstein manifold $M_5$. The form of \eqref{eq:FS3} suggests that there is a similarly universal derivation of this relation from the dual SCFT perspective and it will be most interesting to understand it.

\section{Discussion}\label{sec:conclusion}

We studied infinite families of supersymmetric AdS$_4$ solutions of IIB supergravity arising from D3-branes at a tip of a CY cone over a SE manifold. The Janus solutions are interpreted as holographic duals of interfaces in the 4d $\mathcal{N}=1$ SCFTs associated with the SE manifolds which preserve 3d $\mathcal{N}=1$ superconformal symmetry. When the SE manifold is $S^5$ the Janus solution described above reduces to the one studied in \cite{DHoker:2006vfr,Clark:2005te,Suh:2011xc}. Therefore, it is natural to expect that for a general SE manifold the Janus configurations are similar to the $\mathcal{N}=1$ interfaces in $\mathcal{N}=4$ SYM studied in \cite{DHoker:2006qeo}. It is desirable to investigate further this construction with QFT methods.  

The J-fold solutions should be dual to 3d $\mathcal{N}=1$ SCFTs and it will be most interesting to understand these theories. One possible strategy is to look for a generalization of the construction in \cite{Assel:2018vtq} where the 3d $\mathcal{N}=4$ $T[U(N)]$ theory of Gaitto-Witten \cite{Gaiotto:2008sd,Gaiotto:2008ak} accompanied by appropriate gauging of the flavor symmetries is used to construct the 3d $\mathcal{N}=1$ SCFTs. As in \cite{Assel:2018vtq} the integer $n$ in \eqref{eq:nIIB} is perhaps dual to the Chern-Simons level of the gauge theory. The low amount of supersymmetry makes this system both very interesting and challenging to study. 

Finally we note that it is natural to ask whether there are similar Janus and J-fold solutions of IIB supergravity with $\mathcal{N}=2$ supersymmetry. We will present some explicit examples of such backgrounds in \cite{BGPSvM}.

\section*{Acknowledgements}
We are grateful to S. Pufu and N. Warner for useful discussions. NB is supported in part by an Odysseus grant G0F9516N from the FWO. FFG is a Postdoctoral Fellow of the Research Foundation - Flanders. KP is supported in part by DOE grant DE-SC0011687. MS is supported by the National Research Foundation of Korea under the grant NRF-2019R1I1A1A01060811. The work of JvM is supported by a doctoral fellowship from the Research Foundation - Flanders (FWO). NB, FFG and JvM are also supported by the KU Leuven C1 grant ZKD1118 C16/16/005.

\bibliography{JfoldSE5}

\end{document}